\documentclass[12pt]{article}

\usepackage{latexsym}
\newif\ifams\amsfalse
\amstrue \ifams
 \usepackage{amsfonts}
\fi
\newcommand{\bea}{\begin{eqnarray}}
\newcommand{\eea}{\end{eqnarray}}
\newcommand{\be}{\begin{equation}}
\newcommand{\ee}{\end{equation}}

\ifams
 \def\RR{\mathbb R} \else
 \def\RR{\mbox{\rm$\mbox{I}\!\mbox{R}$}}
\fi

\begin{document}
\title{
\begin{flushright}
\begin{small}
 Alberta Thy 20-99\\
 October 1999 \\
\end{small}
\end{flushright}
\vspace{1.cm}
Holographic Counterterm Actions and Anomalies for Asymptotic AdS
and Flat Spaces}
\author{Jeongwon Ho\\
\small \vspace{- 3mm} Theoretical Physics Institute, Department of Physics,\\
\small \vspace{- 3mm} University of Alberta, Edmonton, Canada T6G 2J1\\
\small E-mail: {\tt jwho@phys.ualberta.ca}}
\date{}
\maketitle

\begin{abstract}
Counterterm actions are constructed along the ADM formalism.
It is shown that the counterterm action can be intrinsically written
in terms of intrinsic boundary geometry. Using the expression of
counterterm action, we obtain a general form of the counterterm action
available for any $d$-dimensional spherical boundary.
In the description, we also derive {\it arbitrary} dimensional holographic
conformal anomaly. It is also shown that counterterm actions for AF
spaces can be obtained from the AdS description just as taking the limit of
$\ell \rightarrow \infty$. An asymptotically flat spacetime with
non-spherical boundary is speculated. In the example, additional
counterterms to eliminate (leading) divergent terms due to deviation
of boundary from round sphere are imagined by observing (4-dimensional)
holographic anomaly proportional to $\Box R$. Argument of the
deceptive-like anomaly is given by comparing with the holographic
description of 5-dimensional Kerr-AdS spacetime.
\end{abstract}
\bigskip
\newpage 
\section{Introduction}
There has been a typical problem to define a gravitational action
suffering from divergence in a non-compact space. In spite that several
prescriptions within the concept of reference space have been suggested
so far \cite{gibbons}\cite{brown}\cite{hawking}, those are flawed by
the fact although the divergences could be eliminated by choosing an
appropriate reference space, it is impossible to embed a boundary
with an arbitrary geometry. Another drawback of the reference space
method is that different reference spaces are needed for different
boundary geometries, so that one cannot define relative energies
in a consistent manner.

Recently, a prominent prescription has been suggested \cite{bal} in the
context of AdS/CFT correspondence \cite{mal}\cite{witten}\cite{gubser},
which could be understood as a realization of the holographic principle
\cite{thooft}\cite{susskind2}. According to the correspondence, UV
divergences of quantum field theory living on a boundary of AdS space are
derived from IR divergences of the bulk theory (UV-IR connection
\cite{susskind}). So, the bulk action could be
regularized by adding local counterterms \cite{witten}\cite{henn}.
On asymptotic AdS spaces, this approach gives an elegant expression of
counterterm action in the form of the expansion for AdS radius $\ell$
\cite{bal}\cite{hyun}\cite{emparan}\cite{kra}
\bea
\label{cotactads}
{\tilde S} &=& -\frac{1}{8\pi G}\int_{\partial X}d^dx\sqrt{-g_0} \left\{
\frac{d-1}{\ell} + \frac{\ell}{2(d-2)}R \right.
\nonumber \\
&& \left. + \frac{\ell^3}{2(d-2)^2(d-4)}\left(
R_{ab}R^{ab} - \frac{d}{4(d-1)}R^2 \right) +
 \cdots \right\}.
\eea

In case of even dimensional boundary, however, one encounters
a logarithmic divergent term in evaluating the bulk action functional.
In order to obtain a finite action if we take the counterterm action
involving this log term, it would cause problematic results in calculating
a boundary stress energy tensor \cite{emparan}. Even though the logarithmic
divergence embarrasses to obtain a finite regularized action, it provides
a remarkable consistency check of the AdS/CFT correspondence
\cite{witten}\cite{henn}. In other words, because a conformal anomaly for
$d$-dimensional conformal field theory in coupling to background gravity
comes from logarithmic UV divergences \cite{birrell}, evaluation of the
conformal anomaly in this scheme becomes a nontrivial check of the AdS/CFT
correspondence. (For holographic conformal anomaly for the dilaton coupled
conformal field theory, see Ref.\cite{odintsov}).

The counterterm action of Eq.(\ref{cotactads}) has been also constructed
from the Gauss-Codazzi equations through an iterative process \cite{kra}.
In the Ref.\cite{kra}, counterterm actions for asymptotically flat (AF)
spaces have been also investigated. However, the procedure adapted for
AdS spaces could not be simply generalized on AF descriptions because of
mathematical difficulty due to non-linearity of the Gauss-Codazzi equations.
Taking an alternative approach, they obtained a counterterm action for AF
spaces with $S^{d-n}\times \RR^n$ boundary geometries
\be
\label{cotactaf}
\tilde{S}= -\frac{1}{8\pi G} \int_{\partial X} d^d x
\sqrt{-g_0} \sqrt{\frac{R^3}{R^2 - R_{ab}R^{ab}}}.
\ee
Very recently, a different prescription to construct the counterterm
action has been suggested in Ref.\cite{solo}. In the prescription, a length
dimensional parameter analogue to the radius of AdS space was defined,
so that the counterterm actions for asymptotically flat and AdS spaces
are consistently constructed in the expansion for the new length parameter.

In this paper, we introduce another method to construct the counterterm
actions in (\ref{cotactads}) and (\ref{cotactaf}). In this construction,
we take the ADM formalism and show that the counterterm action can be
intrinsically written by the terms of intrinsic boundary geometry.
Using our new expression for counterterm action, we obtain a general form
of the counterterm action available for any $d$-dimensional spherical
boundary. In the description, we also derive {\it arbitrary} dimensional
holographic conformal anomaly. It is also shown that the counterterm action
for AF spaces can be obtained from the AdS description just as taking the
limit of $\ell \rightarrow \infty$.

On the other hand, a counterterm action for AF space with nontrivial
boundary geometry is examined. Our counter example is the $D$-dimensional
generalization of the Kerr metric \cite{myers} setting the mass parameter
to zero. It is the metric of an AF space in spheroidal coordinates. This
example has been considered in Ref.\cite{kra}. The authors have
shown that for $d > 6$ the counterterm action in (\ref{cotactaf}) based on
round sphere boundary does not eliminate all divergent terms.
In Ref.\cite{solo}, it has been shown that leading order of divergent terms
due to deviation from round sphere could be canceled by introducing
an additional counterterm action whose a form is similar with the
counterterms of squared boundary curvature in Eq.(\ref{cotactads}).
Here, we shall show that the additional counterterms can be conjectured in
somewhat interesting scheme as well: Our derivation of the counterterm
action does not use the full Einstein equations. Instead only normal-normal
projection equation, which is obtained by projecting the Einstein equations on
the boundary in normal directions, is used. In case of simple boundary
geometry (round sphere), other equations, tangential-tangential and
tangential-normal projections, are not crucial, because they become trivial
or dummy on the procedure. However, in case of nontrivial boundary
geometry, the full Einstein's equations must be used. Our observation is
that taking only the normal-normal projection equation,
we obtain additional divergent terms in boundary action value (BAV),
which are logarithmic. Thus, the conformal invariance of regularized
action (RA) would be broken by a conformal anomaly. However,
we shall show that for an example of $d=4$ the anomaly is proportional
to $\Box R$. That is, additional counterterms proportional to the squared
boundary curvature can be added on the counterterm action in (\ref{cotactaf})
and the conformal invariance would be recovered. We also briefly discuss
about these aspects comparing with AdS descriptions.

Our paper is organized as follows; In Sect.2, counterterm
action is constructed in the ADM formulation.
Examples for asymptotic AdS spaces are considered in Sect.3;
A counterterm action for asymptotic AdS space with $S^d$ boundary is
constructed. In this example, arbitrary dimensional conformal anomaly
is obtained. In Sect.4, relationship between counterterm actions for
asymptotic AdS and flat spaces is discussed. For an AF space in spheroidal
coordinates, the divergent terms are evaluated. Additional counterterms
to eliminate divergent terms due to deviation form round sphere are
conjectured in observation of logarithmic divergence. We also give a
brief discussion about similarities of these aspects and AdS descriptions
in a holographic sense. Discussions and summary are contained in Sect.5.

\section{Holographic Counterterm Actions}
$(d+1)$-dimensional gravitational action with cosmological
constant $ \Lambda =- d(d-1)/(2\ell^2)$ is given by
\begin{equation}
\label{action}
S = \frac{1}{16 \pi G} \int_{X} d^{d+1} x \sqrt{- \hat{G}}
\left( \hat{R} + \frac{d(d-1)}{\ell^2} \right)
-\frac{1}{8 \pi G} \int_{\partial X}d^d x \sqrt{-g} \Theta ,
\end{equation}
where $g_{ab}$ is boundary metric and $\Theta $ is the trace of
extrinsic curvature of $d$-dimensional timelike boundary $\partial X$
defined by $\Theta_{ab} = - g_a^\mu \nabla_\mu n_b $. $\nabla $
denotes the covariant derivative on $(d+1)$-dimensional manifold
$X$ and $n^\mu $ is an outward unit normal to the boundary $\partial X$.
The boundary term in Eq.(\ref{action}), so called Gibbons-Hawking
term, is required for well defined variational principle.

Our purpose is to add another proper surface integral to the
action in (\ref{action}), so that the action becomes finite in the
limit that the boundary is taken to infinity. According to the counterterm
subtraction approach, the additional surface integral
must be written in terms of intrinsic boundary geometry.
For the procedure, we take the ADM formulation as a guide line for
construction of the counterterm action. As it will be seen in the following,
the ADM formulation guarantees for the counterterm action to be written
in terms of intrinsic boundary geometry.

To rewrite the action (\ref{action}) in a canonical form,
we first take a metric given by
\begin{equation}
\label{metric}
\hat{G}_{\mu\nu}dx^{\mu} dx^{\nu} = N^2d\rho^2 + g_{ab}dx^a dx^b,
\end{equation}
where $N^2 =N^2(\rho )$ and $g=g(\rho, x^a)$. In this coordinate
system, unit normal to the boundary is given by
$n_\mu = N \delta^\rho_\mu $. Then, following the standard ADM procedure,
canonical form of the action (\ref{action}) becomes
\begin{equation}
\label{canaction}
S =\int_{X} d^{d+1}x (\pi^{ab}g^{\prime}_{ab} - N{\cal H}_\rho)
 \equiv \int_X d^{d+1}x {\cal L},
\end{equation}
where $\pi^{ab}= \delta {\cal L}/\delta g_{ab}^{\prime}$
is the momentum density conjugate to $g_{ab}$ and $\prime$
denotes the derivative of $\rho$. `Hamiltonian' density
${\cal H}_\rho$ is given by
\begin{equation}
\label{hamil}
{\cal H}_\rho = \frac{16 \pi G}{\sqrt{-g}}
\left( \frac{\pi^2}{d-1} - \pi_{ab}\pi^{ab} \right)
- \frac{\sqrt{-g}}{16 \pi G} \left( R + \frac{d(d-1)}{\ell^2} \right),
\end{equation}
where $R$ is $d$-dimensional scalar curvature of the boundary.
The equation ${\cal H}_\rho = 0$ generates reparametrization of space
coordinate $\rho$. In fact, this equation one of the Gauss-Codazzi
equations that is defined by projecting the Einstein equations on the
boundary in normal directions.

Using the constraint equation ${\cal H}_\rho = 0$, the BAV
evaluated from the action (\ref{action}) on the boundary
$\rho =\rho_0$ is given by a simple form as
\bea
\label{beact}
S_{cl}&=& \int_{\partial X} d^d x \left\{
\frac{1}{8\pi G}\int^{\rho_0} d\rho
N\sqrt{-g} \left( R + \frac{d(d-1)}{\ell^2} \right) \right\}
\nonumber \\
&\equiv & \int_{\partial X} d^d x A(x^a; \rho_0).
\eea
So, according to the counterterm subtraction approach, regularized
action, $S_{RA}$, is defined by
\be
\label{effact}
S_{RA} \equiv S - \tilde{S},
\ee
where the counterterm action $\tilde{S}$ is given by
\be
\label{count}
\tilde{S} =  - \int_{\partial X} d^d x Div \left(A(x^a;\rho_0)\right)
\equiv - \int_{\partial X} d^d x \sqrt{-g_0}\bar{A}^{div}(x^a;\rho_0),
\ee
where $Div$ means to pick divergent terms after
$\rho$-integration and $g_0$ is the induced metric on the boundary.

What a counterterm action is a coordinate invariant functional of intrinsic
boundary geometry is an important requirement in the counterterm subtraction
method. This is because it is the only way to eliminate divergence of a
gravitational action without disturbing the equations of motion or the
symmetries \cite{kra}. In Eq.(\ref{count}), the counterterm action (functional
$A(x^a;\rho_0)$) is explicitly given in terms of intrinsic boundary geometry.
(The `lapse' function $N$ can be absorbed in the space coordinate $\rho$ by a
coordinate redefinition.) In fact, the divergent terms in BAV is originated
from the Gibbons-Hawking term, which is the surface integral of the extrinsic
curvature, as well as from the bulk part. In the procedure, the extrinsic
curvature term is canceled by a term extracted from the bulk part, and the
divergent structure of the BAV in (\ref{beact}) is determined by the terms
originated from bulk part that are expressed in terms of the intrinsic
boundary geometry.

On the other hand, it must be also noted that in the above procedure,
we have not used the full Einstein's equations in obtaining the counterterm
action, but only the constraint equation, ${\cal H}_\rho = 0$, has been used.
In the following, we shall show that in the case of simple boundary
geometry (round sphere), others in Gauss-Codazzi equations become trivial.
Moreover, in the case of nontrivial boundary geometry, this scheme leads
us to somewhat interesting observation. It will be presented in section 4.

\section{AdS Space and Holographic Anomaly}
The counterterm action for asymptotic AdS spaces in (\ref{cotactads}) is
useful for various boundary geometries. However, evaluation of counterterm
actions for higher dimensional boundaries is not manageable for its
mathematical difficulty. In the expression of counterterm action given in
(\ref{count}), we consider a simple but important example, Euclidean AdS
space with $S^d$ boundary and obtain a general form of the counterterm action
available for any $d$-dimensional boundary.

The Euclidean AdS space with $S^d$ boundary is described by the line
element
\begin{equation}
\label{eclads}
\hat{G}_{\mu\nu}dx^\mu dx^\nu = \left( 1 + \frac{r^2}{\ell^2} \right)^{-1}dr^2
+ r^2 d\Omega^2_d.
\end{equation}
The functional $A$ in Eq.(\ref{beact}) for the metric (\ref{eclads}) becomes
\begin{eqnarray}
\label{aftn1}
A(x^a;r_0) &=& \frac{1}{8\pi G} \int^{r_0} dr \sqrt{\gamma_d}
r^d \left(1+ \frac{r^2}{\ell^2} \right)^{-1/2} \left(R +
\frac{d(d-1)}{\ell^2} \right)
\nonumber \\
&=& - \frac{(d(d-1))^{(d+2)/2}}{16\pi G \ell} \sqrt{\gamma_d}
\int^{R_0} dR R^{-(d+2)/2}
\left(1 + \frac{\ell^2 R}{d(d-1)} \right)^{1/2},
\end{eqnarray}
where $R_0$ denotes the scalar curvature on the boundary and $\gamma_d $
is the metric of $d$-dimensional unit sphere. In the second line
of Eq.(\ref{aftn1}), $d(d-1)/r^2 =R$ was used.
After some algebraic calculation, we obtain
\bea
\label{aftn1even}
A(x^a;r_0) &=& - \frac{(d(d-1))^{(d+2)/2}}{16\pi G \ell} \sqrt{\gamma_d}
\left(
\frac{2}{d(d-1)} \sqrt{1+ \frac{\ell^2 R}{d(d-1)}} \right.
\\
&\times & \left. \left\{
- \frac{d-1}{R^{d/2}} + \sum^{(d-2)/2}_{k=1} \left[
\left(- \frac{\ell^2}{d(d-1)} \right)^k \prod^k_{m=1} \left(
\frac{d-2m+1}{d-2m} \right) R^{-(d-2k)/2} \right] \right\} \right.
\nonumber \\
&-&\left. \frac{1}{d} \left( - \frac{\ell^2}{d(d-1)} \right)^{d/2}
\prod^{(d-2)/2}_{k=1} \left(\frac{d-2k-1}{d-2k} \right)
\ln{\frac{\sqrt{1+\ell^2 R/(d(d-1))} -1}{\sqrt{1+ \ell^2 R/(d(d-1))} +1}}
\right)
\nonumber
\eea
in even of $d$ and
\bea
\label{aftn1odd}
A(x^a;r_0) &=&  \frac{d(d(d-1))^{d/2}}{8\pi G \ell} \sqrt{\gamma_d}
\left(1+ \frac{\ell^2 R}{d(d-1)}\right)^{3/2}
\nonumber \\
&& \times \left(- \frac{\ell^2}{d(d-1)} \right)^{(d-5)/2} 
\sum^{(d-3)/2}_{k=0} \prod^k_{m=0} \left(
\frac{d-2m-1}{d-2m} \right)
\eea
in odd of $d$. (After the Eq.(\ref{aftn1even}), we dropped the subscript `0' of
the scalar curvature for simplicity.) Then, an {\it arbitrary} dimensional
counterterm action for AdS spaces with $S^d$ boundary is given
by a polynomial in the boundary scalar curvature $R$ as follows
\bea
\label{ecladscnt}
\tilde{S} &=& \frac{1}{8\pi G}\int_{\partial X}d^dx\sqrt{g_0} \left(
\frac{d-1}{\ell} + \frac{\ell}{2(d-2)}R - \frac{\ell^3}{8d(d-1)(d-4)}R^2
\right.
\nonumber \\
&&\left. + \frac{\ell^5}{16(d(d-1))^2(d-6)}R^3 + \cdots \right),
\eea
where the terms in the parenthesis of Eq.(\ref{ecladscnt}) are terminated by
\begin{equation}
\label{terme}
\frac{1}{2}(-1)^{(d+2)/2}
\prod^{(d+2)/2}_{k=1} \left(\frac{2k-3}{2k}\right)
\frac{\ell^{d+1}}{(d(d-1))^{d/2}}R^{(d+2)/2},
\end{equation}
in the case of $d=even$, and
\begin{equation}
\label{termo}
(-1)^{(d+1)/2}
\prod^{(d-1)/2}_{k=1}\left( \frac{2k-3}{2k}\right)
\frac{\ell^{d-2}}{(d(d-1))^{(d-3)/2}}R^{(d-1)/2},
\end{equation}
in odd $d$ case. Using a relation $R_{ab}R^{ab} =R^2/d $,
it can be shown that the counterterm action in (\ref{ecladscnt}) is equivalent
to the Eq.(\ref{cotactads}). That is, the counterterm action
in (\ref{cotactads}) can be written by a polynomial in the boundary scalar
curvature $R$ terminated by the terms given in (\ref{terme}) or (\ref{termo})
for AdS spaces with $S^d$ boundary. 

On the other hand, the counterterm action for even dimensional boundary
in (\ref{ecladscnt}) fails on eliminating all divergent terms appearing
in the BAV. Instead, the RA contains a logarithmic divergent term
\be
\label{logterm1}
\frac{1}{16 \pi G} \int d^dx \sqrt{g_0}(-1)^{d/2}
\prod^{d/2}_{k=1} \left(\frac{2k-3}{2k}\right)
\frac{\ell^{d-1}}{(d(d-1))^{(d-2)/2}}R^{d/2} \ln{R}.
\ee
It has been already understood in the context of the AdS/CFT correspondence
\cite{witten}\cite{henn}; The regularization of BAV by introducing local
counterterms may break conformal invariance and RA is left with a logarithmic
divergent term. According to this prescription, considering a scale
transformation $\delta r = r \delta \epsilon$ for an infinitesimal constant
parameter $\delta \epsilon$, the holographic conformal anomaly, ${\cal A}$,
for which dual CFT is coupled to the background gravity with $S^d$ boundary
is given by
\be
\label{anomaly1}
{\cal A} = -\frac{1}{8\pi G}(-1)^{d/2}
\prod^{d/2}_{k=1} \left(\frac{2k-3}{2k}\right)
\frac{\ell^{d-1}}{(d(d-1))^{(d-2)/2}}R^{d/2}.
\ee
The conformal anomaly in arbitrary dimensions has been given in geometric
description \cite{deser}. Restricting the CFT in background $S^d$ geometry,
Eq.(\ref{anomaly1}) is an alternative expression of the conformal anomaly
in arbitrary dimensions.

For $S^2$ boundary, the Eq.(\ref{anomaly1}) recovers well known
result
\begin{equation}
\label{anod2}
{\cal A}_{d=2} = -\frac{\ell}{16 \pi G}R.
\end{equation}
Comparing the $(1+1)$-dimensional anomaly on a surface of radius $\ell$,
$-1/(8\pi G \ell) = -c/(12\pi \ell^2)$, the central
charge $c$ becomes $3\ell/(2G)$.
From the Eq.(\ref{anomaly1}) for $d=4$, we find that the conformal anomaly
agrees with that of Ref.\cite{henn}
\be
\label{anod4}
{\cal A}_{d=4} = \frac{\ell^3}{768 \pi G}R^2 = \frac{\ell^3}{8 \pi G}
\left(-\frac{1}{8}R_{ab}R^{ab}+ \frac{1}{24}R^2 \right).
\ee
The conformal anomaly for ${\cal N}=4$ super
Yang-Mill theory on $S^4$ is $3N^2/(8\pi^2 \ell^4)$.
Comparing with the anomaly on this boundary from the Eq.(\ref{anod4}),
$3/(16\pi G \ell)$, we obtain the expected result
\be
\label{rankn}
N^2= \frac{\pi \ell^3}{2 G},
\ee
where $N$ is the rank of the gauge group of the dual ${\cal N}=4$
supersymmetric $d=4$ $SU(N)$ YM theory.
At last, it can be seen that for six dimensional boundary, the anomaly
in (\ref{anomaly1}) is equivalent to that given in \cite{henn}
\bea
\label{anod6}
{\cal A}_{d=6}&=& - \frac{\ell^5}{115200 \pi G} R^3
\nonumber \\
 &=& -\frac{1}{16 \pi G} \left(\frac{\ell^5}{64}\right) \left(
 \frac{1}{2} R R_{ab} R^{ab}
 - \frac{3}{15} R^3
 - R^{ab} R_{acbd} R^{cd} \right. \nonumber \\
 &&  \left.
 + \frac{1}{5} R^{ab} D_a D_b  R
 - \frac{1}{2} R^{ab} \Box R_{ab}
 + \frac{1}{20} R \Box R \right).
\eea
In fact, since we are concerned about $S^6$ boundary, the terms in third line
including derivatives vanish. On the other hand, Eq.(\ref{anod6}) can be
verified by considering the central charge of $N$ coincident M5-branes in
the large $N$ limit. It has been shown that the central charge is proportional
to $N^3$ \cite{gubser2}. So, the anomaly on $S^6$ boundary with radius
$\ell$, $15/(64 \pi G \ell)$, is proportional to $N^3/(\pi^4 \ell^6)$.
Thus, we find \cite{mal}
\be
\label{centn}
N^3 \sim \frac{\pi^3 \ell^5}{G}.
\ee

Before ending of this section, it is useful on the next section
to consider another Euclidean AdS space with
different boundary geometry, $S^{d-1} \times S^1$,
\be
\label{schads}
 ds^2 = 
 \left( 1+ \frac{r^2}{\ell^2} \right) d\tau^2
+ \left( 1+ \frac{r^2}{\ell^2} \right)^{-1}dr^2
 + r^2 d\Omega_{d-1}^2~.
\ee
For the $S^{d-1} \times S^1$ boundary, the functional $A$ in (\ref{beact})
becomes
\be
\label{aftn2}
A(x^a;r_0) = \frac{\sqrt{\gamma_{d-1}}}{8\pi G}
\left(\frac{d-1}{\ell^2} \right)
\left( \frac{(d-1)(d-2)}{R} \right)^{d/2} \left(
1+ \frac{\ell^2 R}{(d-1)(d-2)} \right).
\ee
Since all terms in expanding of Eq.(\ref{aftn2}) are divergent for $d>2$,
the counterterm action is just the negative of $S_{cl}$ in (\ref{beact})
\cite{kra}
\be
\label{schadscnt}
 \tilde{S}= \frac{1}{8\pi G}\int_{\partial X} d^d x \sqrt{g_0}
\left(\frac{d-1}{\ell}
\right)\left(1+ \frac{\ell^2 R}{(d-1)(d-2)} \right)^{1/2}.
\ee
It can be shown that using $R_{ab}R^{ab} = R^2/(d-1)$ and expanding for
$\ell$, the counterterm action in (\ref{schadscnt}) is equivalent to
Eq.(\ref{cotactads}). However, it must be noted that while the counterterm
action for $S^d$ boundary is given by a finite sum of the series in
(\ref{cotactads}), for $S^{d-1} \times S^1$ boundary it is given by an
infinite sum. As mentioned in Ref.\cite{kra}, in the process the divergent
factors $1/(d-4)$, $1/(d-6)$, $\cdots$ in (\ref{cotactads}) are canceled.
Thus, while conformal invariance of the RA for $S^d$ boundary is broken
by the anomaly in (\ref{anomaly1}), for the $S^{d-1} \times S^1$ boundary
it is still conformal invariant.

\section{AF Space and Holographic Anomaly}
Now, consider counterterm actions for asymptotic flat spaces.
In Ref.\cite{kra}, it has been shown that the counterterm action for AF
spaces is not be able to obtain on taking a limit of $\ell \rightarrow \infty$
in the procedure adapted for AdS spaces, an iteration process using the
Gauss-Codazzi equations, because of mathematical difficulty. However,
in our procedure, those are simply obtained by taking the $\ell \rightarrow
\infty$ limit on the functional $A$'s
\be
\label{afcnt1}
\tilde{S}
= -\frac{1}{8\pi G} \int_{\partial X} d^d x
\sqrt{-g_0} \sqrt{\frac{dR}{d-1}}
\ee
in (\ref{aftn1even}) and (\ref{aftn1odd}), and
\be
\label{afcnt2}
\tilde{S}
= -\frac{1}{8\pi G} \int_{\partial X} d^d x
\sqrt{-g_0} \sqrt{\frac{d-1}{d-2}R}
\ee
in (\ref{aftn2}). In Eqs.(\ref{afcnt1}) and (\ref{afcnt2}), the
counterterm actions were written in the Lorentzian signature.
Given in Ref.\cite{kra}, the counterterm actions in (\ref{afcnt1}) and
(\ref{afcnt2}) can be written by
\be
\label{afcntgen}
\tilde{S}= -\frac{1}{8\pi G} \int_{\partial X} d^d x
\sqrt{-g_0} \sqrt{\frac{R^3}{R^2 - R_{ab}R^{ab}}}.
\ee
In fact, the expression of counterterm action for AF spaces in
(\ref{afcntgen}) is more general than those in (\ref{afcnt1}) and
(\ref{afcnt2}). Because it is available for the AF spaces with $S^{d-n}
\times \RR^n$ boundary geometries described by the metric
\be
\label{afmetric}
\hat{G}_{\mu\nu}dx^\mu dx^\nu = (-dt^2 + dx^2_1+ \cdots + dx^2_{n-1})
+ dr^2 + r^2d\Omega^2_{d-n}.
\ee
For AF spaces described by the metric (\ref{afmetric}), the functional $A$
becomes
\be
\label{aftnaf}
A(x^a; \rho_0) = \frac{1}{8\pi G}\int^{\rho_0} d\rho N\sqrt{-g} R,
\ee
and then the counterterm action is
\bea
\label{afcnt3}
 \tilde{S} &=& -\frac{1}{8\pi G}\int_{\partial X} d^d x \sqrt{-g_0}
\sqrt{\frac{d-n}{d-n-1}R}
\nonumber \\
&=&  -\frac{1}{8\pi G} \int_{\partial X} d^d x
\sqrt{-g_0} \sqrt{\frac{R^3}{R^2 - R_{ab}R^{ab}}}.
\eea

Up to now, we have considered counterterm actions for manifolds with simple
boundary geometry (round sphere). Now, we speculate divergences of BAV due to
deviation of boundary from the round sphere. In Ref.\cite{kra}, an AF space
in spheroidal coordinates given by
\be
\label{kerrmz}
\hat{G}_{\mu\nu} dx^\mu dx^\nu= -dt^2
+ {\rho^2 \over r^2 + a^2} dr^2 + \rho^2 d\theta^2
+ \sin^2{\theta} (r^2 +a^2) d\phi^2
+ r^2 \cos^2{\theta} d\Omega_{d-3}^2
\ee
was investigated. This space can be obtained by setting the mass to zero in
higher dimensional Kerr metric \cite{myers}
\bea
\label{kerrbh}
\hat{G}_{\mu\nu}dx^\mu dx^\nu &=&
-\frac{\Delta}{\rho^2}\left(dt - a\sin^2{\theta} d\phi \right)^2
+\frac{\sin^2{\theta}}{\rho^2}\left(adt - (r^2 + a^2)d\phi \right)^2
\nonumber \\
&& +\frac{\rho^2}{\Delta}dr^2 + \rho^2 d\theta^2
+ r^2 \cos^2{\theta} d\Omega_{d-3}^2,
\eea
where $\rho^2 =r^2 + a^2 \cos^2{\theta}$, $\Delta = r^2 -2mGr^{4-d} +a^2$, and
$m$ and $a$ are the black hole mass and the angular momentum per
unit mass, respectively. It is important that the metric in (\ref{kerrmz})
does not describe the asymptotic spacetime of the Kerr black hole in
(\ref{kerrbh}). Because, in the process, one is to meet a naked singularity.
It is just the flat spacetime metric of $n=1$ in Eq.(\ref{afmetric})
written in spheroidal coordinates. 

The functional $A(x^a;r_0)$ for the metric (\ref{kerrmz}) becomes
\bea
\label{aftnkerrmz}
A(x^a;r_0) &=& \frac{\sqrt{\gamma_{d-3}}}{8 \pi G}
\int^{r_0}dr r^{d-3} \left[
\frac{2a^2\left(
(d-3)\sin^2{\theta} - \cos^2{\theta} \right)}{
r^2 + a^2\cos^2{\theta}}
\right.
\nonumber \\
&&\left. +(d-1)(d-2) + (d-3)(d-4)\frac{a^2}{r^2} 
\right]\sin{\theta}\cos^{d-3}{\theta}
\nonumber \\
&=& \frac{\sqrt{\gamma_{d-3}}}{8 \pi G}r_0^{d-2} \left[ (d-1) +\left(
d-3 + \frac{2((d-3)\sin^2{\theta} - \cos^2{\theta})}{d-4} \right)
\frac{a^2}{r_0^2} \right.
\nonumber \\
&&\left.-\frac{2\cos^2{\theta}((d-3)\sin^2{\theta} - \cos^2{\theta})}{d-6}
\frac{a^4}{r_0^4} + \cdots\right]\sin{\theta}
\cos^{d-3}{\theta},
\eea
where the divergence terms in the bracket are terminated by
\be
\label{termtereven}
2a^2\left(
(d-3)\sin^2{\theta} - \cos^2{\theta} \right)
(- a^2\cos^2{\theta})^{(d-4)/2}r_0^{-(d-2)} \ln{r_0}
\ee
in even $d$, and
\be
\label{termterodd}
2a^2\left(
(d-3)\sin^2{\theta} - \cos^2{\theta} \right)
(- a^2\cos^2{\theta})^{(d-5)/2} r_0^{-(d-3)}
\ee
in odd $d$, respectively. In the above calculation, the $d$-dimensional scalar
curvature $R$ is
\bea
\label{curkerr}
R &=& \frac{2a^2\left((d-3)\sin^2{\theta} - \cos^2{\theta}\right)}{
(r^2 + a^2\cos^2{\theta})^2} + \frac{(d-3)(d-4)}{r^2\cos^2{\theta}}
\nonumber \\
&&+\frac{\left(2(2d-5) -(d-3)(d-4)\tan^2{\theta}\right)}{
r^2 + a^2\cos^2{\theta}}.
\eea

The appearance of logarithmic divergence in Eq.(\ref{termtereven}) looks
like strange. But, this is an artificial effect due to the fact that
we have not used yet the remaining Einstein's equations that are
normal-tangential and tangential-tangential projections.
In fact, rewriting the terms in the bracket of Eq.(\ref{aftnkerrmz}) as
\bea
\label{diffterm}
&&(d-1) + \left(d-2-\cos^2{\theta} + \frac{d\sin^2{\theta} - 2}{d-4} \right)
\frac{a^2}{r_0^2}
\nonumber \\
&&+\left( (\cos^4{\theta} -\cos^2{\theta}) 
+ \frac{\cos^2{\theta}
(\cos^2{\theta} - d\sin^2{\theta})}{d-6} \right)\frac{a^4}{r_0^4}
+ \cdot \cdot \cdot,
\eea
comparing with the calculation of Ref.\cite{kra}, some additional
terms appear, $r_0^{-2}/(d-4),~r_0^{-4}/(d-6),\cdots$, which may cause the
logarithmic divergent term of (\ref{termtereven}). After substituting
the remaining equations into the action, those additional terms and the
logarithmic divergence would vanish.

However, this logarithmic divergence seems to tell us something more. 
Following the procedure taken in the AdS descriptions,
the logarithmic divergence breaks conformal invariance of the
RA within a conformal anomaly
\begin{equation}
\label{anomaly2}
{\cal A}^{flat} = \frac{r_0^{d-1}}{4 \pi G}\left[ \frac{
a^2(-a^2 \cos^2{\theta})^{(d-4)/2}\left(
(d-3)\sin^2{\theta} - \cos^2{\theta} \right)}{
r_0^{2(d-2)}\sqrt{(r^2_0 + a^2\cos^2{\theta})
(r^2_0 + a^2)}} \right].
\end{equation}
For a strict argument, consider 4-dimensional boundary whose the conformal
anomaly, ${\cal A}_{d=4}^{flat}$, is
\be
\label{afanod4}
{\cal A}_{d=4}^{flat}= \frac{r_0^3}{4 \pi G}(\sin^2{\theta} - \cos^2{\theta})
\left(\frac{
a^2}{r_0^6}\right)\left(1+ \frac{a^2 \cos^2{\theta}}{r_0^2}\right)^{-1/2}
\left(1+ \frac{a^2}{r_0^2} \right)^{-1/2}.
\ee
Then we find that up to leading order the anomaly is proportional to $\Box R$
\be
\label{afanod4sim}
{\cal A}_{d=4}^{flat}\sim\frac{r_0^3}{4 \pi G} \left(- \frac{1}{40} \Box R
+ {\cal O}\left(\frac{a^4}{r_0^8} \right) \right).
\ee
As well known in the dual field theory on curved boundary, the
4-dimensional anomaly has an ambiguity that $\Box R$ term can be added to
the anomaly with an undetermined coefficient. This corresponds to the choice
of different schemes for regularizing the field theory. In our case,
this comes under different choices for the counterterm action.
So, we can add additional counterterms given by
\be
\label{addcotact1}
\Delta \tilde{S} \sim r^3_0 \int_{\partial X} d^4 x \sqrt{-g_0}
\left( a E + b C_{abcd}C^{abcd} + c R^2 \right),
\ee
where $E$ is the Euler invariant $E = R_{abcd}R^{abcd} - 4 R_{ab}R^{ab}
+ R^2$ and $C^{abcd}$ is the Weyl tensor. Taking the
coefficients as $b= -a$, $c=0$, then the additional counterterms become
\be
\label{addcotact2}
\Delta \tilde{S} \sim -2a r^3_0 \int_{\partial X} d^4 x \sqrt{-g_0}
\left( R_{ab}R^{ab} - \frac{1}{3}R^2 \right).
\ee
The additional counterterm action in (\ref{addcotact2}) is just that
observed by Solodukhin in Ref.\cite{solo}; Those counterterms cancel
the divergent terms which are caused by deviation of boundary from
round sphere. It should be also noted that in spite that the counterterm
actions in Eqs.(\ref{afcnt2}) and (\ref{afcntgen}) based on round
sphere boundary exactly cancel divergent terms of the gravitational
action for $d<6$, the additional counterterm action in (\ref{addcotact2})
is not proportional to the cubed boundary curvature $r^3_0 \int_{\partial X}
R^3$ but $r^3_0 \int_{\partial X} R^2$. This is because the first
correction (i.e., small deviation from round sphere) identically vanishes
\cite{solo}, i.e., leading terms of $R_{ab}R^{ab}$ and $R^2$ are canceled
each other in the brace of Eq.(\ref{addcotact2}).

\section{Summary and Discussions}
The counterterm subtracting method to define a finite gravitational action
on non-compact spacetime has been speculated.
It has been shown that using the ADM formalism, the counterterm action
could be explicitly written in terms of the intrinsic boundary geometry.
On the other hand, using the form of counterterm action, we have obtained
an expression of counterterm action available for {\it arbitrary} dimensional
AdS spaces with $S^d$ boundary geometry. Moreover, from this expression
the arbitrary dimensional conformal anomaly has been driven. Our additional
observation is that the counterterm action for AF spaces can be obtained
as taking the limit of $\ell \rightarrow \infty$ in the procedure adapted for
AdS spaces. 

Another interesting observation in the description for counterterm action
developed in this paper has been given in the example of AF space with
nontrivial boundary geometry. It has been shown that the additional
counterterms to eliminate (leading) divergent terms due to deviation of
boundary from round sphere, which was suggested by Solodukhin \cite{solo},
can be imagined from appearance of logarithmic divergence in BAV and
perspective of the corresponding anomaly proportional to $\Box R$. It seems
that it is due to a deceptive procedure as skipping over tangential-tangential
and tangential-normal projections of Einstein's equations in calculation of
BAV. In fact, simply using the full Einstein's equations, then we would obtain
a BAV without the logarithmic divergent term. However, it appears that
there is something in the holographic sense. For the 5-dimensional Kerr-AdS
spacetime with the boundary of 4-dimensional rotating Einstein universe,
the trace of stress tensor does not vanish, but the evaluated BAV does not
contain a corresponding logarithmic divergence \cite{adel}. As mentioned in
the Ref.\cite{adel}, the corresponding logarithmic divergence of the BAV
should not be present for a spacetime which can be written locally
as a product. It has been also observed that the conformal anomaly, which
corresponds to the CFT anomaly, is also proportional to the $\Box R$
term. In addition, it can be shown \cite{ho} that taking the limit of
$\ell \rightarrow \infty$ on the Kerr-AdS description, the conformal anomaly
becomes the Eq.(\ref{afanod4sim}). These holographic similarities of the
Kerr spacetime with the Kerr-AdS description will be studied in detail
in Ref.\cite{ho}. It is also expected that this study will be of help
to prescribe the problem of construction of the flat-space S-matrix
which has been studied on the large radius limit in the AdS/CFT
correspondence and suffered from non-local holographic mapping
\cite{susskind3}\cite{giddings}.

\vspace{0.2in} {\bf Acknowledgments:} I thank V. Frolov, A. Zelnikov, and
Y. Gusev for helpful discussions. This work was supported
in part by Korea Science and Engineering Foundation and in part by 
National Science and Engineering Research Council of Canada.


\begin{thebibliography}{10}

\bibitem{gibbons}
G.W. Gibbons and S.W. Hawking, Phys. Rev. D{\bf 15} (1977) 2752.

\bibitem{brown}
J.D. Brown and J.W. York Jr, Phys. Rev. D{\bf 47} (1993) 1407.

\bibitem{hawking}
G.T. Horowitz and S.W. Hawking, Class. Quantum Grav. {\bf 13} (1996) 1487.

\bibitem{bal}
V. Balasubramanian and P. Kraus, Commun. Math. Phys. {\bf 208} (1999) 413.

\bibitem{mal}
J. Maldacena, Adv. Theor. Math. Phys. {\bf 02} (1998) 231.

\bibitem{witten}
E. Witten, Adv. Theor. Math. Phys. {\bf 02} (1998) 253.

\bibitem{gubser}
S.S. Gubser, I.R. Klebanov, and A.M. Polyakov, Phys. Lett. B{\bf 428}
(1998) 105.

\bibitem{thooft}
G. 't Hooft, {\it `Dimensional Reduction in Quantum Gravity'}, in
{\it Salamfest} (1993) p.284, gr-qc/9310026.

\bibitem{susskind2}
L. Susskind, J. Math. Phys. {\bf 36} (1995) 6377.

\bibitem{susskind}
L. Susskind and E. Witten, {\it `The Holographic Bound in Anti-de Sitter
Space'}, hep-th/9805114.

\bibitem{henn}
M. Henningson and K. Skenderis, JHEP {\bf 07} (1998) 023.

\bibitem{hyun}
S. Hyun, W.T. Kim, and J. Lee, Phys. Rev. D{\bf 59} (1999) 084020.

\bibitem{emparan}
R. Emparan, C.V. Johnson, and R.C. Myers, Phys. Rev. D{\bf 60} (1999) 104001.

\bibitem{kra}
P. Kraus, F. Larsen, and R. Siebelink, Nucl. Phys. B{\bf 563} (1999) 259.

\bibitem{birrell}
N.D. Birrell and P.C.W. Davies, {\it `Quantum Fields in Curved Space'},
Cambridge University Press, Cambridge (1982).

\bibitem{odintsov}
S. Nojiri and S.D. Odintsov, Phys. Lett. B{\bf 444} (1998) 92.

\bibitem{solo}
S.N. Solodukhin, {\it `How to make the gravitational action on
non-compact space finite'}, hep-th/9909197.

\bibitem{myers}
R.C. Myers and M.J. Perry, Ann. Phys. {\bf 172} (1986) 304.

\bibitem{deser}
S. Deser and A. Schwimmer, Phys. Lett. B{\bf 309} (1993) 279.

\bibitem{gubser2}
S.S. Gubser and I.R. Klebanov, Phys. Lett. B{\bf 413} (1997) 41.

\bibitem{adel}
Adel M. Awad and Clifford V. Johnson, {\it Holographic Stress Tensors
for Kerr-AdS Black Holes}, hep-th/9910040.

\bibitem{ho}
Jeongwon Ho, {\it in preparation}.

\bibitem{susskind3}
L. Susskind, {\it Holography in the Flat Space Limit'}, hep-th/9901079.

\bibitem{giddings}
S.B. Giddings, {\it Flat-Space Scattering and Bulk Locality
in the AdS/CFT Correspondence'}, hep-th/9907129.

\end{thebibliography}
\end{document}